# Large Thermoelectricity via Variable Range Hopping in Chemical Vapor Deposition Grown Single-layer MoS$_2$

*Jing Wu* ‡ [1,2,3], *Hennrik Schmidt* ‡ [1,2], *Amara Kiran Kumar*[4], *Xiangfan Xu*[5], *Goki Eda*[1,2,4], *and Barbaros Özyilmaz*[1,2,3*]

[1] Department of Physics, National University of Singapore, 117542, Singapore

[2] Graphene Research Center, National University of Singapore, 117542, Singapore

[3] NanoCore, National University of Singapore, 117576, Singapore

[4] Department of Chemistry, National University of Singapore, 117542, Singapore

[5] Center for Phononics and Thermal Energy Science, School of Physics Science and Engineering, Tongji University, 200092 Shanghai, China

ABSTRACT: Ultrathin layers of semiconducting molybdenum disulfide ($MoS_2$) offer significant prospects in future electronic and optoelectronic applications. Although an increasing number of experiments bring light into the electronic transport properties of these crystals, their thermoelectric properties are much less known. In particular, thermoelectricity in chemical vapor deposition grown $MoS_2$, which is more practical for wafer-scale applications, still remains unexplored. Here, for the first time, we investigate these properties in grown single layer $MoS_2$. Micro-fabricated heaters and thermometers are used to measure both electrical conductivity and thermopower. Large values of up to ~30 mV/K at room temperature are observed, which are much larger than those observed in other two dimensional crystals and bulk $MoS_2$. The thermopower is strongly dependent on temperature and applied gate voltage with a large enhancement at the vicinity of the conduction band edge. We also show that the Seebeck coefficient follows $S \sim T^{1/3}$ suggesting a two-dimensional variable range hopping mechanism in the system, which is consistent with electrical transport measurements. Our results help to understand the physics behind the electrical and thermal transports in $MoS_2$ and the high thermopower value is of interest to future thermoelectronic research and application.

Two-dimensional crystals have recently attracted much attention due to their exciting properties and also their potential for practical applications (1-4). Especially molybdenum disulfide ($MoS_2$), a member from the group of transition-metal dichalcogenides, holds great promise for spin-electronics and optoelectronics (5-8). Besides the electrical characterization, the understanding of thermal properties such as thermopower (TEP), expressed in the Seebeck coefficient *S*, of $MoS_2$ is important to gain insight on the design of high performance devices with low power consumption. The TEP of a material depends on its band structure and electron and hole asymmetries at the Femi level (9). Therefore it is also an extremely sensitive tool to characterize the electronic structure as well as the density of states (10). This applies even if the contacts to such nanoscaled systems are not perfect for transport characterization because the TEP measurement does not need the induction of an external current.

From the application side, the direct conversion between heat and electricity in thermoelectric materials is considered to be a promising route for power generation to solve the energy shortage problem, and accordingly, the search for highly efficient thermoelectric materials has attracted much attention in material science. Recently, Buscema et. al studied the photothermoelectric properties of mechanically exfoliated $MoS_2$ and observed large TEP in this material, evidencing its potential application on thermopower generation and waste heat harvesting (11). However, a systematic thermoelectric study, which is also important for understanding the charge carrier transport mechanism, is still missing. This is probably due to the fact that the mechanical exfoliation of thin flakes proved to be more complicated compared to, for example, graphene. Recently, the success of high quality growth of large area monolayer $MoS_2$ using Chemical Vapor Deposition (CVD) (12, 13) provides the possibility for the realization of complex large area device concepts oriented towards applications.

Here, we report on electrical and thermopower measurements of single layer CVD $MoS_2$ over a wide temperature range (20-300 K) by employing microfabricated heaters and thermometers. Large TEP of 30 mV/K is observed, which is 2 orders of magnitude larger than that in pristine graphene (14, 15) and other nanoscaled materials (16-19, 21, 22) and also 1 order of magnitude larger than that of bulk $MoS_2$. We further show that the gate modulated TEP is greatly enhanced in the low carrier density region of $MoS_2$ and can be quantitatively related to its electrical conductance in this regime through the Mott relation.

Single layer $MoS_2$ is grown by CVD on a $Si/SiO_2$ substrate following previous studies (23). Afterwards, the $MoS_2$ film is transferred (see supporting information) onto highly doped silicon substrates with 285 nm of oxide layer which acts as a dielectric layer to tune the charge carrier density. Electrodes are fabricated using standard electron-beam lithography followed by thermal evaporation of Cr (2 nm)/Au (50 nm) in such a way that the part which is in contact with the $MoS_2$ consists of pure gold. $XeF_2$ has been used previously to etch MoS2 (24), and the same technique is applied in our experiment to achieve the required device geometry and to electrically decouple the sample from the heater and the remaining areas of grown MoS2 films. A schematic layout of the device, similar to previous TEP studies on one-dimensional Nanowires (16-22) and graphene (14, 15, 25) is shown in Figure 1a. The device consists of a heater and two local thermometers ($R_{hot}$ and $R_{cold}$). The inset in Figure 1b shows an optical image of the device.

It has been shown that vacuum annealing changes the doping of the $MoS_2$ by shifting the Fermi level towards the conduction band and also reduces the contact resistance significantly (26, 27). We apply similar annealing steps and measure the conductance during the annealing, which gives us certain control over the final doping level of the device. After the annealing, a large shift (~80 V) of threshold gate voltage towards negative values was observed and the resistance of the two probe measurement decreases from 1 M$\Omega$ to 50 k$\Omega$ at 60 V back gate voltage (see supporting information).

We obtained the source-drain characteristics of the sample at different gate voltages and temperatures by using the two thermometers as source and drain leads. Figure 1b shows the transfer curves at highest and lowest temperature (additional data in the supporting information). The device exhibits an on-off ratio exceeding $10^4$ at room temperature and $10^6$ at 5 K, the field effect mobilities are 15 $cm^2V^{-1}s^{-1}$ and 55 $cm^2V^{-1}s^{-1}$ respectively. The maximum 2-terminal conductance reaches ~ 20 $\mu$S, which is comparable with previously reported results (27, 28). A clear shift of the threshold gate voltage towards to the n-doping direction can be observed while the temperature increases from 5 K to 300 K (26, 29). At a voltage around 60 V, a crossing of the curves is observed, which is consistent with previous published results (26-28) and gives information about the transition to metallic behavior.

After the electronic characterization, TEP measurements are carried out using a low-frequency Lock-in setup. An AC-current $I_{heater}$ with the frequency $\omega$=13.373 Hz is driven through the heater to create a heat gradient over the sample by Joule heating. The heating power is given by $P=I_{heater}^2 R_{heater}$, in which $R_{heater}$ is the resistance of the heater electrode, and the frequency of the heat gradient $\triangle T$ induced by the heating current is $2\omega$ accordingly. The resulting $2\omega$ TEP voltage drop, $V_{TEP}$, is measured between the two thermometer electrodes ($R_{hot}$ and $R_{cold}$). The heat gradient $\triangle T$ across the relevant area is determined by probing the four-probe resistances of the two thermometers $R_{hot}$ and $R_{cold}$. The TEP of the single layer CVD MoS$_2$ is then given by the Seebeck coefficient $S=-\triangle V_{TEP}/\triangle T$.

Figure 2 shows the back gate dependence of the TEP at different temperatures. The negative values of TEP indicate that electrons are the majority charge carriers, which is consistent with the vicinity to the conduction band. The obtained maximum value for the TEP at 280 K yields 30 mV/K, which is remarkably larger than other investigated low-dimensional materials, such as graphene ($\sim\pm100$ µV/K) (14, 15), Bi$_2$Te$_3$ ($\sim\pm200$ µV/K) (17, 18), semiconducting carbon nanotubes ($\sim-300$ µV/K) (16), Ge-Si core-shell Nanowires (~400 µV/K) (21), and InAs Nanowires ($\sim-5$ mV/K) (20). Note that our TEP values may be even underestimated due to the limitation of measurement setup at the high resistive off-state of the material (30) (see supporting information). Taking this into account, the TEP of single-layer CVD MoS$_2$ is comparable to the previous studies on exfoliated MoS$_2$ (~-100mV/K) (11). Although the ZT value of MoS$_2$ is rather low due to the low conductance, such high TEP values are still interesting for thermoelectronics studies and its potentially applications. The TEP values measured here are also significantly higher than the ones observed in bulk MoS$_2$ (~7mV/K) (31). In the bulk, the charge carrier density cannot be fully tuned by gate voltages. On the other hand, atomically thin layers can be very homogenously gated and therefore will exhibit a more uniform potential landscapes than its bulk counterpart. This allows a precise adjustment of the Fermi level towards the variable-range hopping (VRH) regime, as discussed below, where thermopower is maximized.

In the range of back gate voltages studied, three distinct regimes (I, II, III) can be identified. When the back gate voltage is at high positive values (III), the TEP shows very small values of around 1~10 µV/K, which is comparable to the values of metallic materials (32-34). Together with the conductance data in Figure 1b, this indicates that the system approaches the metallic regime. As the back gate voltage is reduced, the Fermi level is shifted into the bandgap and MoS$_2$ undergoes a transition from metallic to insulating behavior (region II). The carrier density and the conductance decreases as expected for a semiconductor. At the same time, the TEP starts to increase and finally it reaches the maximum value. This is followed by a sharp decrease in TEP as the back gate voltage is swept to higher negative values (region I). This sharp decrease coincides with the "off" state in MoS$_2$ indicating that the TEP closely follows the conductance variations in MoS$_2$.

It should also be noted that the measured TEP values includes not only the contribution from CVD MoS$_2$ but also from the gold electrodes deposited on the MoS$_2$, leading to a TEP of $S_{device}=S_{MoS2}+S_{electrode}$. However $S_{electrode}$ is three to four orders of magnitude smaller compared to the one from MoS$_2$ (11) and it is also expected to give a constant contribution independent of the back gate voltage. Therefore the TEP of the device can be considered as the TEP of the single layer CVD MoS$_2$.

In order to understand the gate dependence of the TEP in region II, one has to have a look at the dominating transport mechanisms of MoS$_2$. The Mott formula, which relates the TEP to the electronic conductance, is given by (35):

$$S = -\frac{\pi^2}{3}\frac{k_B^2}{e}T \times \frac{1}{G}\frac{dG}{dV_g}\frac{dV_g}{dE}|_{E=E_F}$$

where $E_F$, $k_B$ and $e$ are the Fermi energy, Boltzmann constant and electron charge, respectively. From this relationship, the absolute value of the TEP is expected to increase monotonically while the charge carrier density and the conductance G decrease. However, with an increasingly negative back gate voltage, the system will reach the insulating regime. The conductivity at this regime is almost energy independent and hardly changes with the back gate. At the same time, the amount of electrons conducting through the material goes towards zero at the insulating state. In this regime (region I), the TEP measurement is limited by the measurement setup due to high device resistance (30) and the $V_{TEP}$ signal decreases very quickly and exhibiting unreliable values once MoS2 goes to this 'off' state.

In addition to the gate voltage dependence, we investigate the temperature dependence of the TEP. As T increases, the TEP shifts towards band gap. In Figure 3 a and b, we show the conductance and TEP as a function of back gate voltage and temperature. As the temperature increased from 5 K to 300 K, the same shift in conductance and TEP can be observed. In Figure 3c, cross sections of conductance and TEP at a constant temperature of T = 80 K are shown. The TEP starts to increase at the threshold back gate voltage where electrons start conducting and decreases when the conductance around 0.1 µS for all temperatures measured. The TEP peak position and threshold-voltage are compared for different temperatures in Figure 3d. The peak of TEP shifts from -10 V to -56 V as the temperature increases from 20 K to 300 K and follows the same trend of the threshold voltage. After comparing the TEP peak position ($V_{PEAK}$) and threshold-voltage ($V_{TH}$), we find that $V_{PEAK}$-$V_{TH}$ is almost constant. The same trend verifies that the electrical and thermal transports depend on the same mechanism, that is, electrical conduction and thermal power are dominated by electrons and not by phonon-drag effects in the thermoelectric transport.

To have a further understanding of the scattering mechanism on the electrical and thermal transport, we apply the Mott VRH model to fit the temperature dependent conductance at low charge carrier densities (36, 37). This model describes a system which is strongly disordered with the charge carriers hopping between localized states. The VRH relationship between conductance and temperature is given by (38)

$$\sigma = \sigma_0 exp[-(T_0/T)^{1/(d+1)}]$$

in which $\sigma$ is the conductivity, $d$ is the dimensionality of the system and the prefactor is given by $\sigma_0 = AT^{-m}$, with the constants $m$ and $A$. In this case, $d$ equals 2 because of the two-dimensionality of the single layer CVD $MoS_2$. Figure 4a shows the conductance of the device as a function of temperature at different gate voltages from 60 V to -20 V. It can be well fitted by the VRH equation at different temperatures using $m=0.8$, which is in good accordance to previous results on electronic transport in the high resistance regime (29). However, although the values fit very good for low temperatures, the VRH approach does not work as well in the high temperature range (beyond 200 K) due to the fact that thermal excitation of charge carriers becoming more dominant. Because the thermoelectric properties are determined by the charge transport in the system, the TEP is expected to follow similar behavior at low temperatures. The revised TEP by incorporating the Mott formula can be written as (39, 40)

$$S \approx k_B T (\frac{T_M}{T})^{\frac{2}{d+1}} \times \frac{dn(E)}{dE}|_{E=E_F} \propto T^{\frac{d-1}{d+1}}$$

where $T_M$ is the temperature independent Mott coefficient and $n(E)$ is density of states. Using $d=2$, we expect $S \propto T^{1/3}$. In Figure 4b, we plot the TEP values at $V_{BG}-V_{TH}= 20V$ as the function of $T^{1/3}$ with the temperature range of 20-300 K. From the linear fits, the low temperature (T<200 K) dependence agrees very well with the VRH Mott relation expected for the disordered semiconductor. The analysis in the regimes of lower resistivity yields similar results (see supporting information). As long as the system is dominated by VRH, the contribution from phonon-drag effect on TEP is negligible (41, 42). This can be also seen in the close connection between the back gate voltage dependence of the TEP and the electrical conductance as mentioned earlier. These observations can be attributed to the strong localization of the charge carriers since phonon-drag only affects the non-localized ones (43). When the system goes towards higher carrier densities as the back gate voltage increases, the Fermi level moves up and the localized states in the gap start getting filled up (44), resulting in the VRH transport observed in this experiment. Furthermore, as temperature increases, thermally excited charge carriers also start to contribute to the process. This contribution increases very fast with temperature. When the Femi level passes over the localized states and gets close to the conduction band, the contribution from thermally excited charges increases and the VRH will be suppressed as observed in our measurements.

To summarize, electrical and thermal measurements of single layer CVD $MoS_2$ have been performed in a broad temperature range. Remarkably large TEP values, up to 30 mV/K are observed, which is attributed to the tunability and the mechanism of electronic transport of ultrathin $MoS_2$ and give these two dimensional semiconductors potential for thermoelectric applications. From the great enhancement at low carrier density region and the shift with the temperature, we show that at low carrier densities, transport is dominated by the variable-range hopping. This link between TEP and conductance also indicates that TEP measurements can be used to understand both the electrical and thermal properties of this low dimensional material.

**Figure 1**. (a) Schematic of the thermoelectric device. Two electrodes, $R_{hot}$ and $R_{cold}$, contacting with the single layer CVD MoS$_2$ flake, are used as 4-probe thermometers to determine the heat gradient applied by the heating current $I_{heater}$, and also to measure the thermoelectric voltage drop. A back gate voltage $V_{BG}$ is applied to tune the carrier density of the device. (b) Electrical transport characteristics of the single layer CVD MoS$_2$ measured at 5K and 300K. The inset shows an optical image of the contacted device.

**Figure 2**. (a) Thermopower of single layer CVD MoS$_2$ as a function of back gate voltage at different temperatures. The grey shaded region indicates the measurement uncertainty. Three distinct regimes (I, II, III) are identified, shifting with temperature.

**Figure 3**. Conductance (a) and TEP (b) as a function of back gate voltage and temperature. (c) Comparison of electric transport and TEP as at the cross sections at 80K indicated in the upper graphs. (d) TEP peak position and threshold voltage of conductance as a function of temperature.

**Figure 4** (a) VRH behavior of the conductance at different back gate voltages. (b) The TEP values at the gate voltage $V_{BG}$-$V_{TH}$= 20V as a function of temperature. The dotted line shows the fit indicating VRH at lower temperatures.

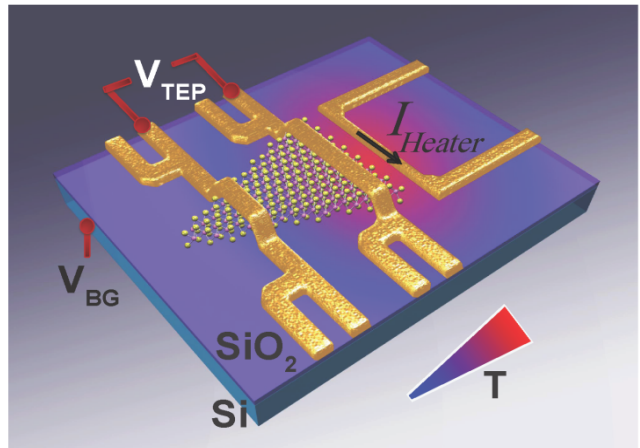

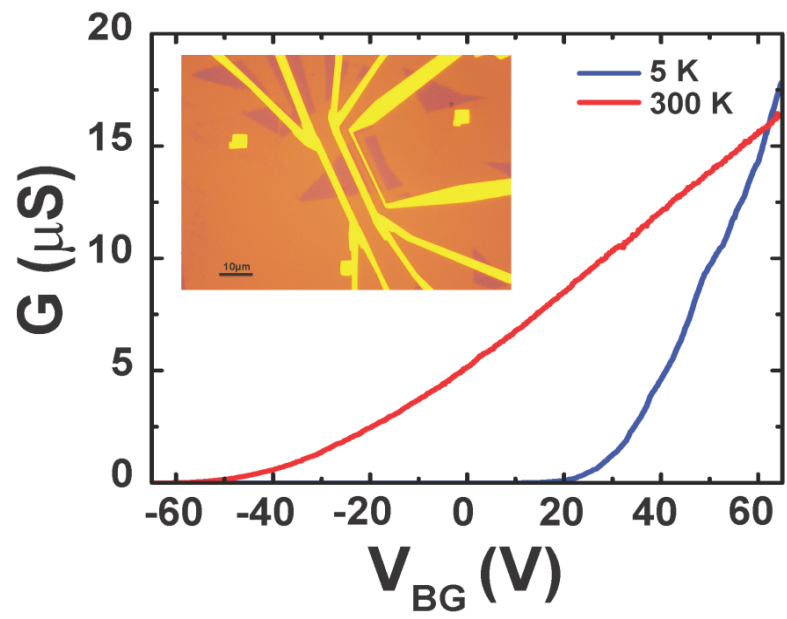

**Fig. 1 Wu. J. et al.**

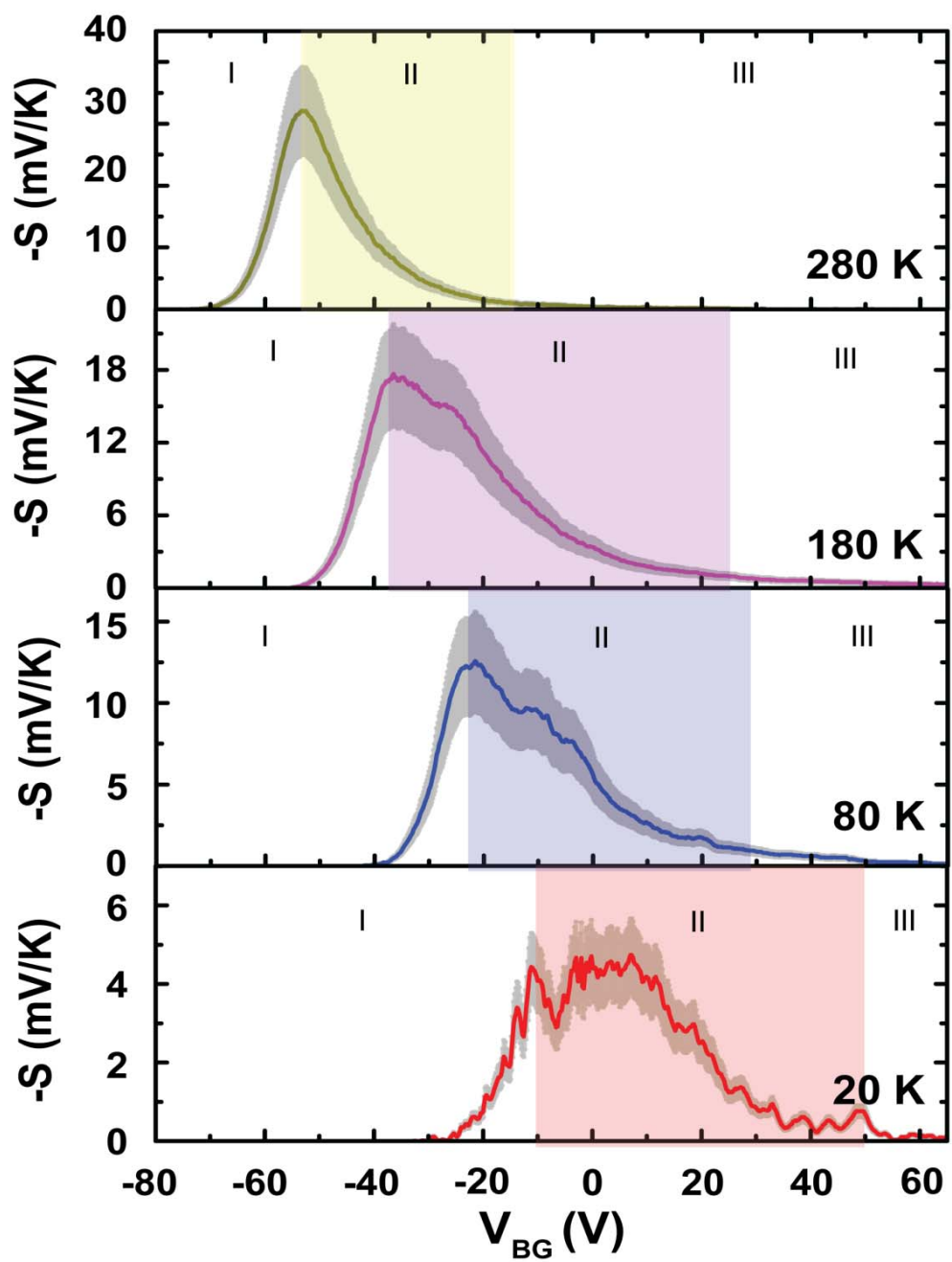

**Fig. 2 Wu. J. et al.**

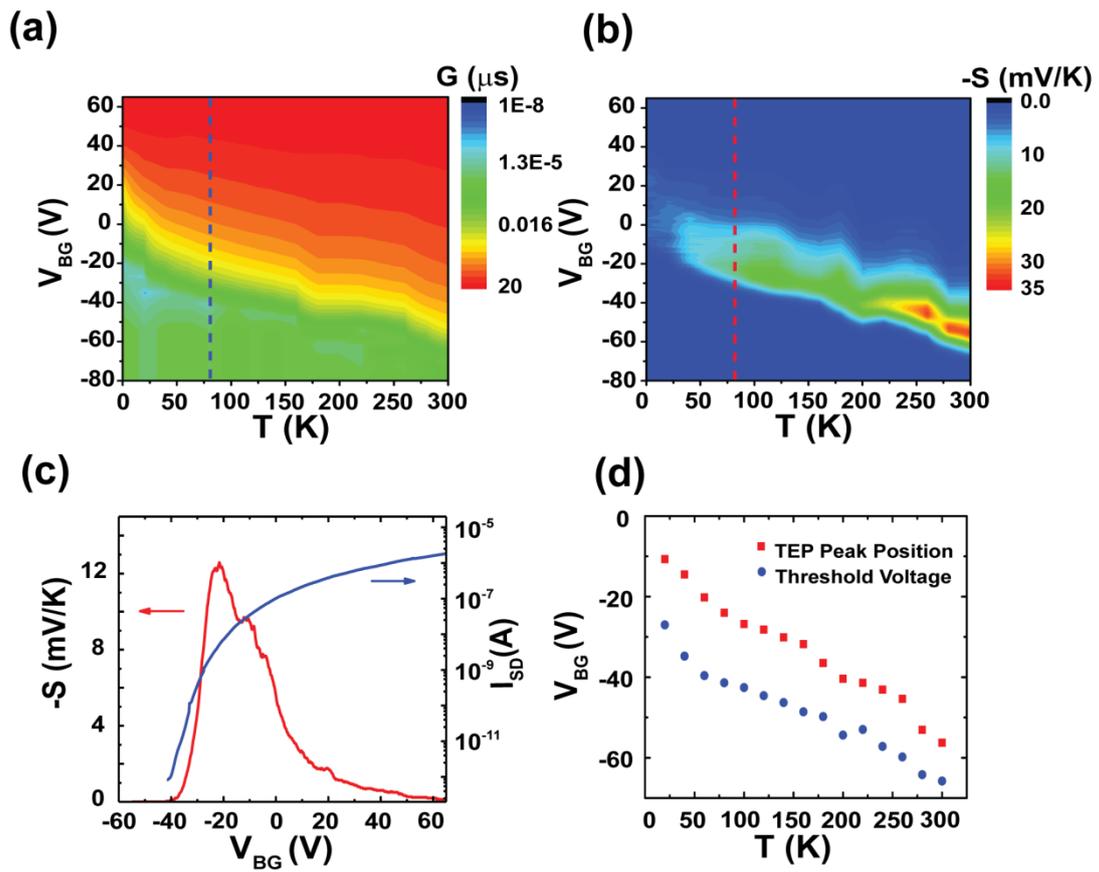

Fig. 3 Wu. J. et al.

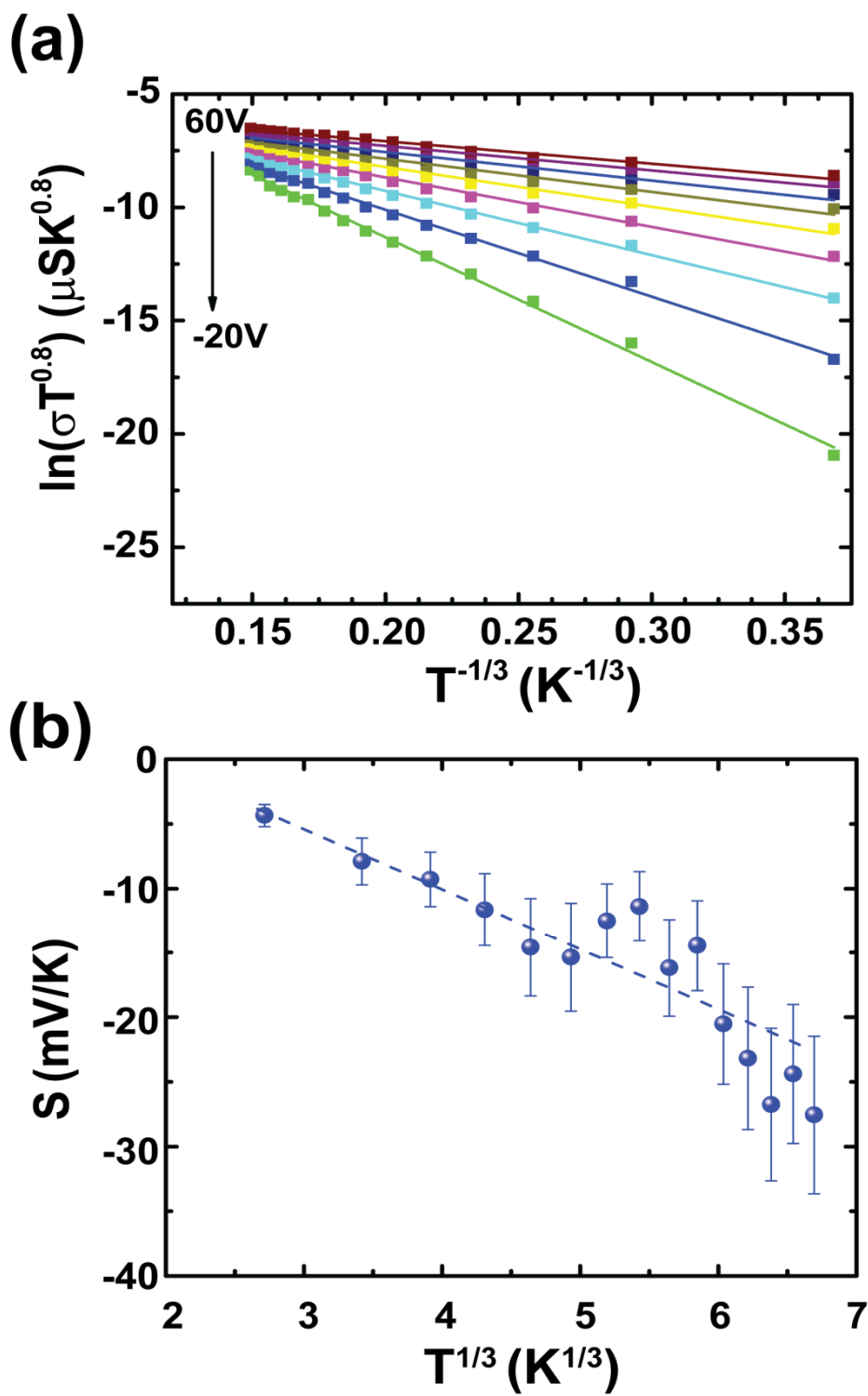

**Fig. 4 Wu. J. et al.**

ASSOCIATED CONTENT

**Supporting Information**: Details on the grows process, sample transfer, contacts, and annealing steps, Raman mapping data, additional transport experiments, on chip thermometer calibration, heat power dependence of the TEP, reproduced high TEP values.

AUTHOR INFORMATION

**Corresponding Author**

\*B. Ö. Email: phyob@nus.edu.sg

**Author Contributions**

B. Ö. supervised the project. J. W. and B. Ö. designed the experiments. J. W. and H. S. performed the experiments. A. K. K. grown the CVD $MoS_2$. All authors carried out the data analysis and discussed the results. J. W., H. S. and B. Ö. wrote the manuscript.

‡These authors contributed equally.

**Funding Sources**

B. Ö. acknowledges support by the NUS Young Investigator Award, the Singapore National Research Foundation Fellowship award (RF2008-07-R-144-000-245-281), the NRF-CRP award (R-144-000-295-281) and the Singapore Millennium

Foundation-NUS Research Horizons award (R-144-001-271-592; R-144-001-271-646).

G.E acknowledges Singapore National Research Foundation for funding the research under NRF Research Fellowship (NRF-NRFF2011-02, NRF-NRFF2012-01)

X.X. acknowledges National Research Foundation China (No. 11304227).

ACKNOWLEDGMENT

The authors would like to thank Jayakumar Balakrishnan, Ivan J. Vera-Marun and Gavin Kok Wai Koon for helpful discussions and comments.

# Large Thermoelectricity via Variable Range Hopping in CVD Grown Single-layer MoS$_2$

*Jing Wu ‡ $^{1,2,3}$, Hennrik Schmidt ‡ $^{1,2}$, Kiran Kumar Amara$^4$, Xiangfan Xu$^5$, Goki Eda$^{1,2,4}$, and Barbaros Özyilmaz$^{1,2,3*}$*

[1] Department of Physics, National University of Singapore, 117542, Singapore

[2] Graphene Research Center, National University of Singapore, 117542, Singapore

[3] NanoCore, National University of Singapore, 117576, Singapore

[4] Department of Chemistry, National University of Singapore, 117542, Singapore

[5] Center for Phononics and Thermal Energy Science, School of Physical Science and Engineering, Tongji University, 200092 Shanghai, China

**Sample preparation:**

**CVD growth:** Single layers of $MoS_2$ were grown by Chemical Vapour deposition in a furnace (MTI) from $MoO_3$ (Alfa Aesar) and S (Sigma Aldrich) precursors. The $Si/SiO_2$ growth substrates were cleaned by sonication in Acetone and IPA for 15-20min and then immersed in piranha solution (1:3 mixtures of $H_2O_2$ and $H_2SO_4$) for 2hrs before the growth was performed. 14mg of $MoO_3$ was spread uniformly over 2-3cm in an alumina crucible and $Si/SiO_2$ growth substrates were places upside down on the crucible. Sulfur powder was placed upstream in 30sccm of nitrogen gas flow. After placing the precursors in the furnace, the quartz tube was evacuated to $10^{-4}$ mbar and then flushed with nitrogen to remove the contamination from atmospheric gases. The quartz tube was initially heated to 105°C in 5min and held at that temperature for 30min to remove the moisture and then heated to 555°C in 30min, 700°C in 10min, held at 700°C for 5min and cooled down naturally to room temperature without any feedback.

**Transfer:** $MoS_2$ films were transferred to another clean $Si/SiO_2$ chip before measurements. 400 nm PMMA (A5 950K) was spin coated on the top of the CVD $MoS_2$. This PMMA layer was used as a supporting layer for the MoS2 transfer. After spin coating, the chip with $MoS_2$ was putted into a KOH solvent to etch away the $SiO_2$ layer. After the etching process, the PMMA layer will floats on the top of the solvent. At the same time, $MoS_2$ also leaves the grown substrate and float with the PMMA. Few clean steps in DI water followed the etching before we transferred the $MoS_2$ on to new substrates. PMMA with $MoS_2$ was scooped from DI water by another clean $Si/SiO_2$ chip, then bake at 180 °C on the hot plate for 2 minutes. The PMMA layer was washed away by acetone later.

**Contacts:** Two steps of electron-beam lithography and thermal evaporation have been used for making the contacts of the devices. We evaporate Cr (2nm)/Au (50nm) for the big outside contacts and use pure Au (50nm) for the small electrode which contact with the $MoS_2$ and give reproducible good contacts.

**Etch**: Since our CVD grown $MoS_2$ is continuously in a relative large area, some parts have to be etched to obtain an isolated device structure. After the contact deposition, we employed an etch method to isolate the target $MoS_2$ from the surrounding crystals. From Figure S1, one can find that the etch process is very clean and defined.

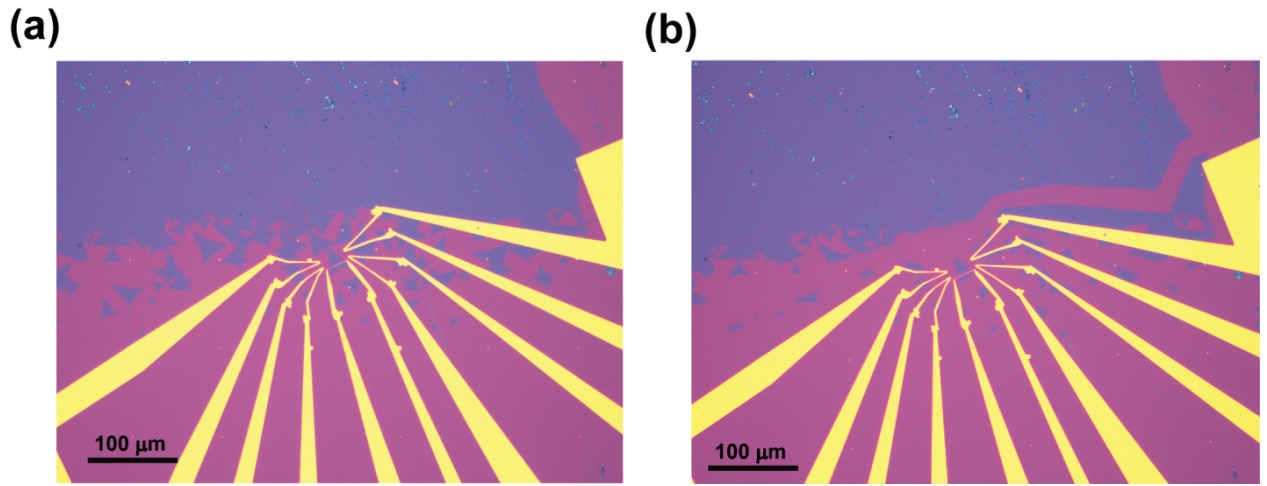

**Figure S1.** (a) (b) optical image of the device before and after etching.

**Anneal process:** Before our measurements, the samples were annealed in $N_2$ environment at 200 °C to pre-clean the device by removing residues resulting from the fabrication process. After that, sample was loaded into the cryostat and in situ annealing was performed in vacuum at 400 K for several hours. From the two probe $I_{SD}$-$V_{BG}$ measurement (Figure S2 (a)), a very large shift (~80V) of threshold gate-voltage towards negative values was observed after the annealing process. We also monitor the conductance of the sample during the second in situ annealing in the cryostat. Figure S2 (b) shows that the conductance at 0V back gate voltage increases continuously when the device is annealed. We use this information to stop the annealing process when the sample is in the medium conduction regime. The temperature curve shown in Figure S2 (b) is obtained from the on chip thermometer.

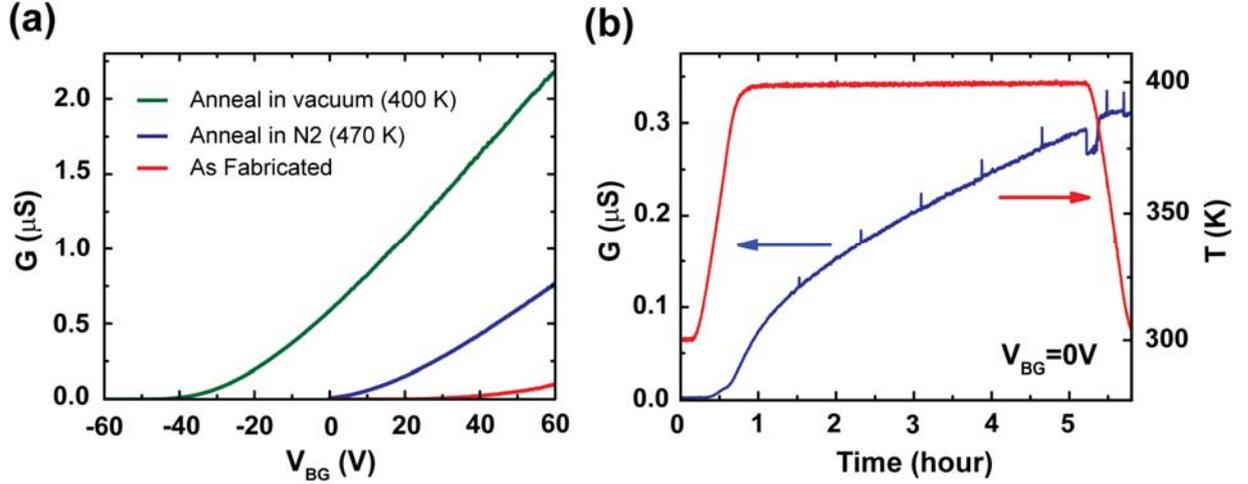

**Figure S2.** (a) Conductance vs $V_{BG}$ before and after anneal. (b) Conductance at 0V back gate voltage during the in situ annealing process at 400K.

## Sample characterization:

Raman spectroscopy is used for MoS$_2$ characterization. Figure S3 (a) is the single spectrum of our CVD MoS$_2$ sample. The distance between $E'_{2g}$ peak and $A_{1g}$ peak is 21 cm$^{-1}$ which confirm the single layer CVD MoS$_2$ of our sample according to literature (1). Figure S3 (b) shows the 2D mapping of the $A_{1g}$ peak of the device (Figure S3 (b)). Compared to the optical image, the dark lines A, B and C in the mapping picture is the thermometer $R_c$, $R_h$ and heater respectively.

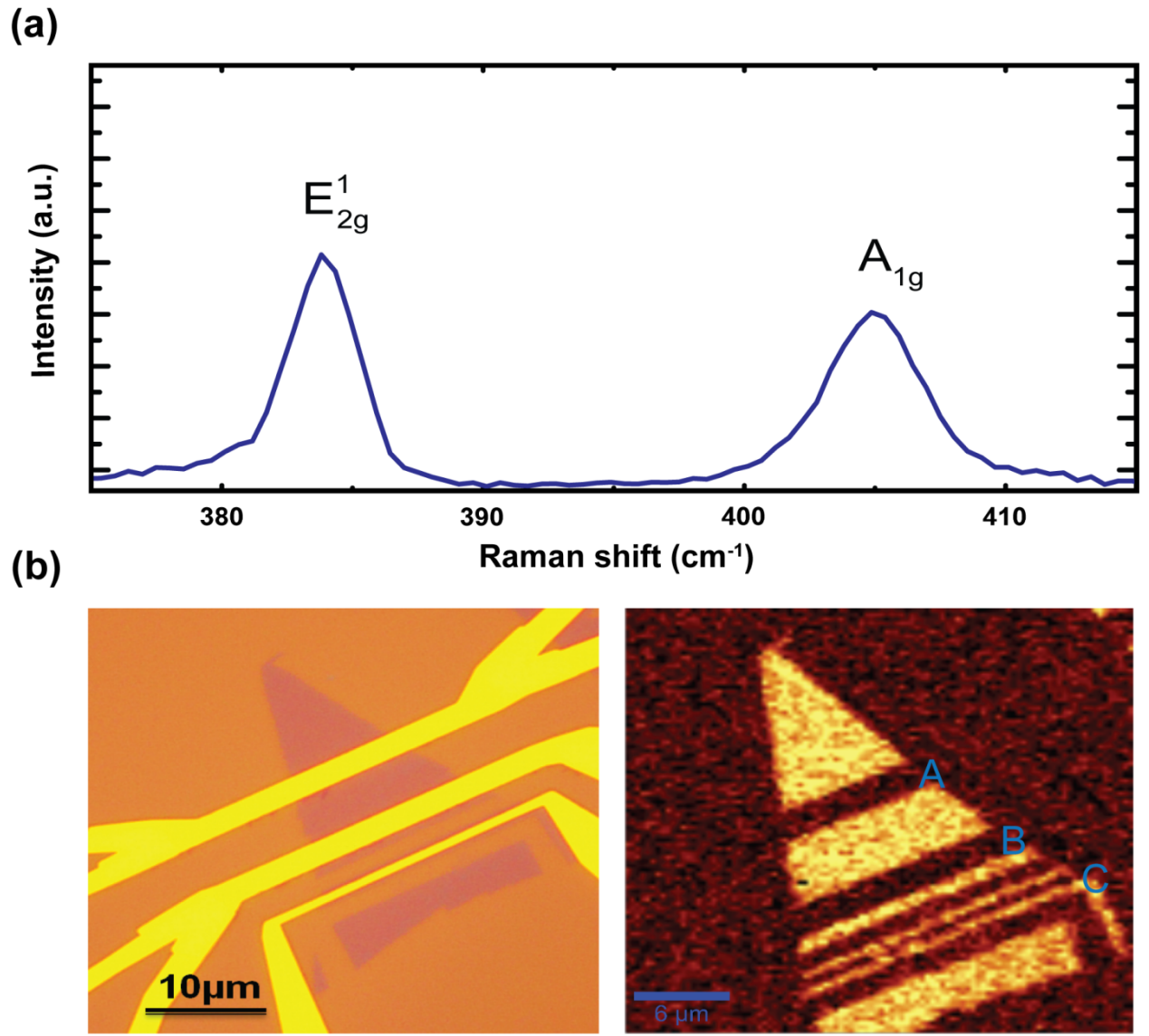

**Figure S3.** (a) Single Raman spectrum of the CVD MoS$_2$ device. (b) Optical image and 2D Raman mapping of A$_{1g}$ peak of the device.

After annealing, we applied source-drain characteristics of the sample by using the two thermometers. Figure S4 shows the I$_{SD}$ as the function of back gate in different temperatures.

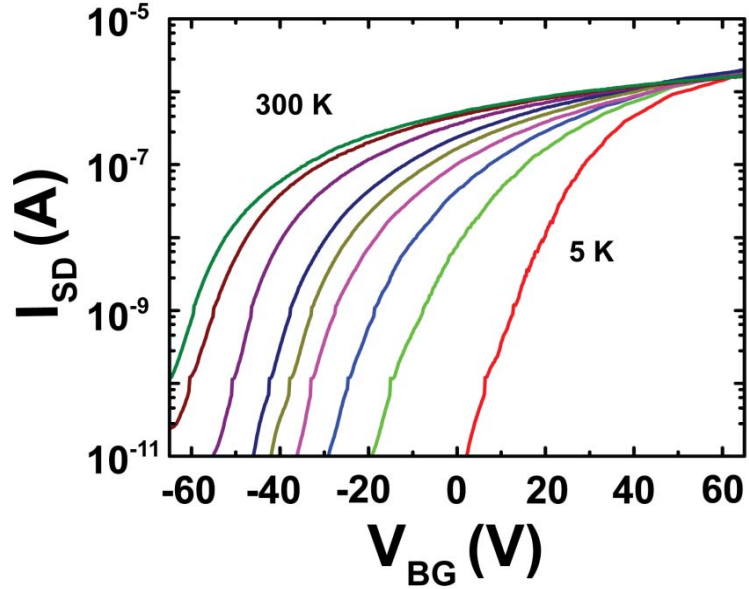

**Figure S4.** $I_{SD}$ vs $V_{BG}$ at different temperatures: from 5K to 300K.

## Temperature calibration:

Temperature gradient $\triangle T$ is determined by probing the four-probe resistance of two thermometers $R_{hot}$ and $R_{cold}$. We first calibrate the temperature coefficient of $R_{hot}$ and $R_{cold}$ by changing the environment temperature in the measurement system. The temperature of the environment was monitored by the thermometers built-in on PPMS sample socket. Figure S5 (a) and (b) show the resistance of $R_{hot}$ and $R_{cold}$ as the function of temperature after thermal equilibrium was reached. The linear results serve as basic calibration curves of thermometry. Such temperature calibration was performed for temperatures between 20 K to 300 K. Since the resistance of the thermometers saturates at low temperature, the temperature measurements will be inaccurate in this regime.

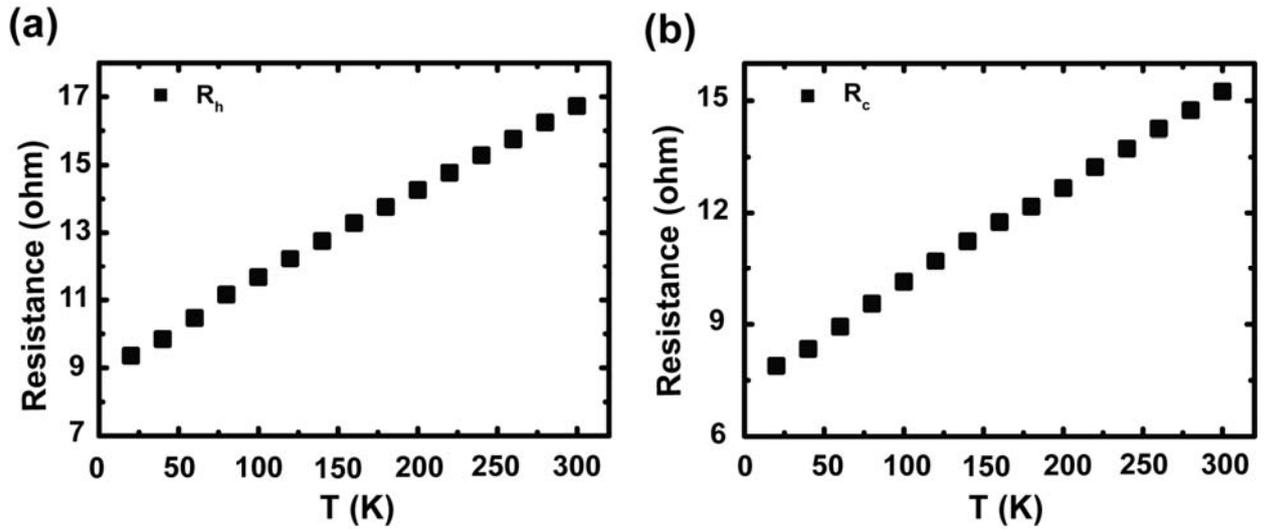

**Figure S5.** (a), (b): Resistance calibration of thermometers $R_h$ and $R_{cold}$.

After the calibration of the thermometers, a heating current to induce the heat gradient was applied. At the same time, the resistance of $R_{hot}$ and $R_{cold}$ where measured. When the environment temperature and thermometer resistance stabilized, we compare the values of the $R_{hot}$ and $R_{cold}$ to the calibration curve to obtain the temperature increase at these two thermometers, $\triangle T_{hot}$ and $\triangle T_{cold}$. The temperature gradient of the sample $\triangle T = \triangle T_{hot} - \triangle T_{cold}$ can be calculated accordingly (Figure S6).

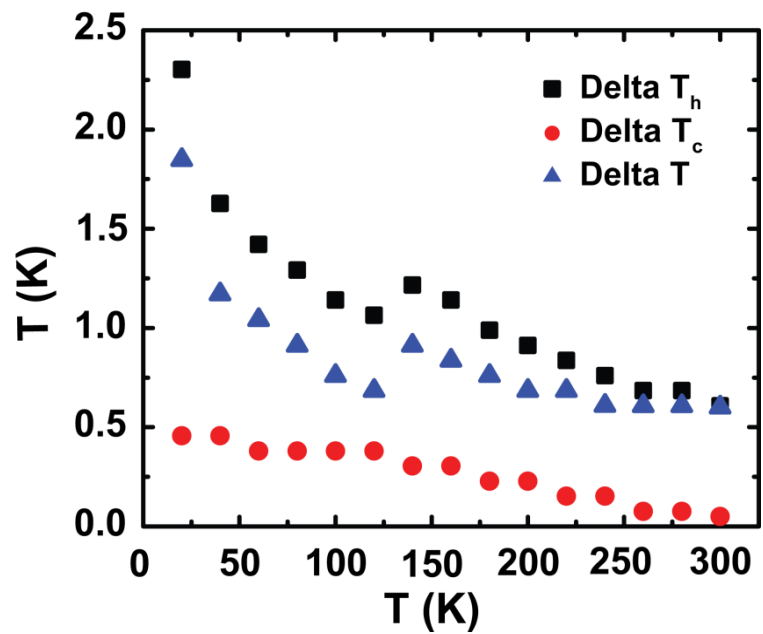

**Figure S6.** The black squares and red circles indicate the temperature changes on two thermometers at different temperatures. The blue triangles show the temperature gradient induced by the heater at different temperatures.

## Measurement Setup:

Figure S7 shows the details of our measurement setup. A Keithley 6430 is used as DC source meter for the transfer curve characterization and conductance measurement (a). The circuit diagram of the AC configuration using Lock-in amplifiers SR830 for the TEP measurement is shown in (b).

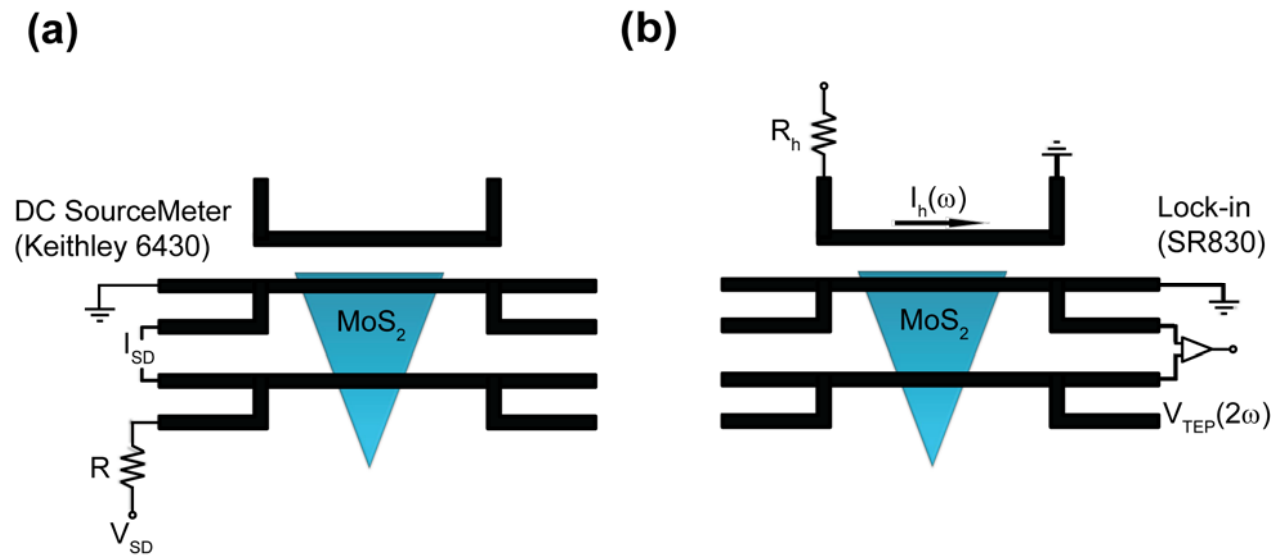

**Figure S7. (a)** and **(b)** show the measurement setup of DC and AC configuration

## Additional thermopower results:

Figure S8 (a) shows the TEP at 20 K cryostat temperature as a function of different heating power. The temperature difference of the samples is proportional to the heating power, which can be seen in the linear relationship between the TEP and heating power.

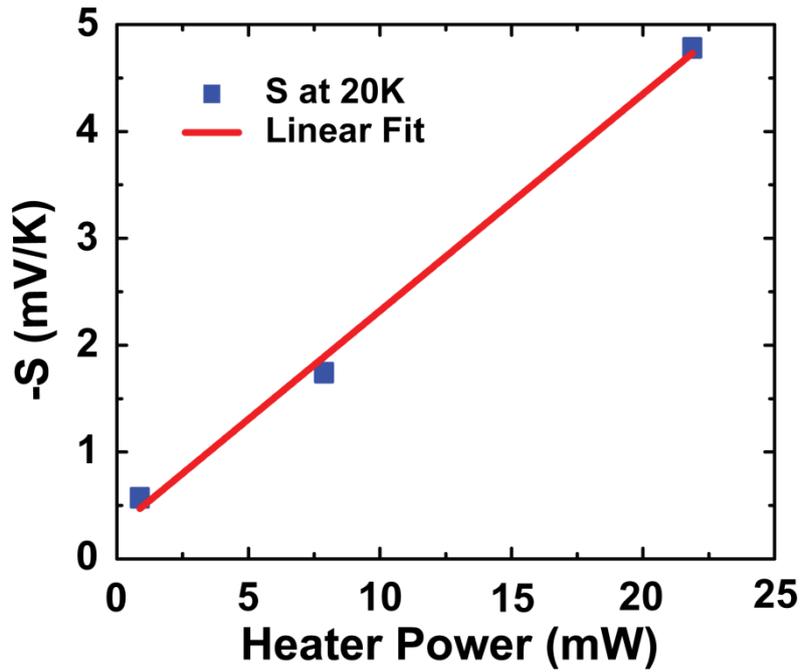

**Figure S8.** TEP results at 20K with different heating power.

When $MoS_2$ is gated towards the insulating region, the resistance of the device goes beyond the input impedance of our measurement setup (10 MΩ) and the thermoelectric voltage measurement of the device becomes inaccurate. The resulting voltage drop will be underestimated due to this limitation. To make sure the temperature dependence discussed in Figure 4b is not influenced by this, similar analysis can be performed in the region, where the resistance of the device is well below 10 MΩ.

Figure S9. (a) to (d) show the TEP values at gate voltage $V_{BG}-V_{TH}$= 30V, 40V, 45V, 50V. The resistance of the device in these cases is 5 MΩ, 1.3 MΩ, 700 kΩ and 500 kΩ, which is smaller than the input impedance of the measurement setup. The temperature ranges from 20K to 120K in which the contribution from thermal excitation is still not too strong. The resulting fits are consistent with the once shown in the main text and verify the proposed VRH Mott relation.

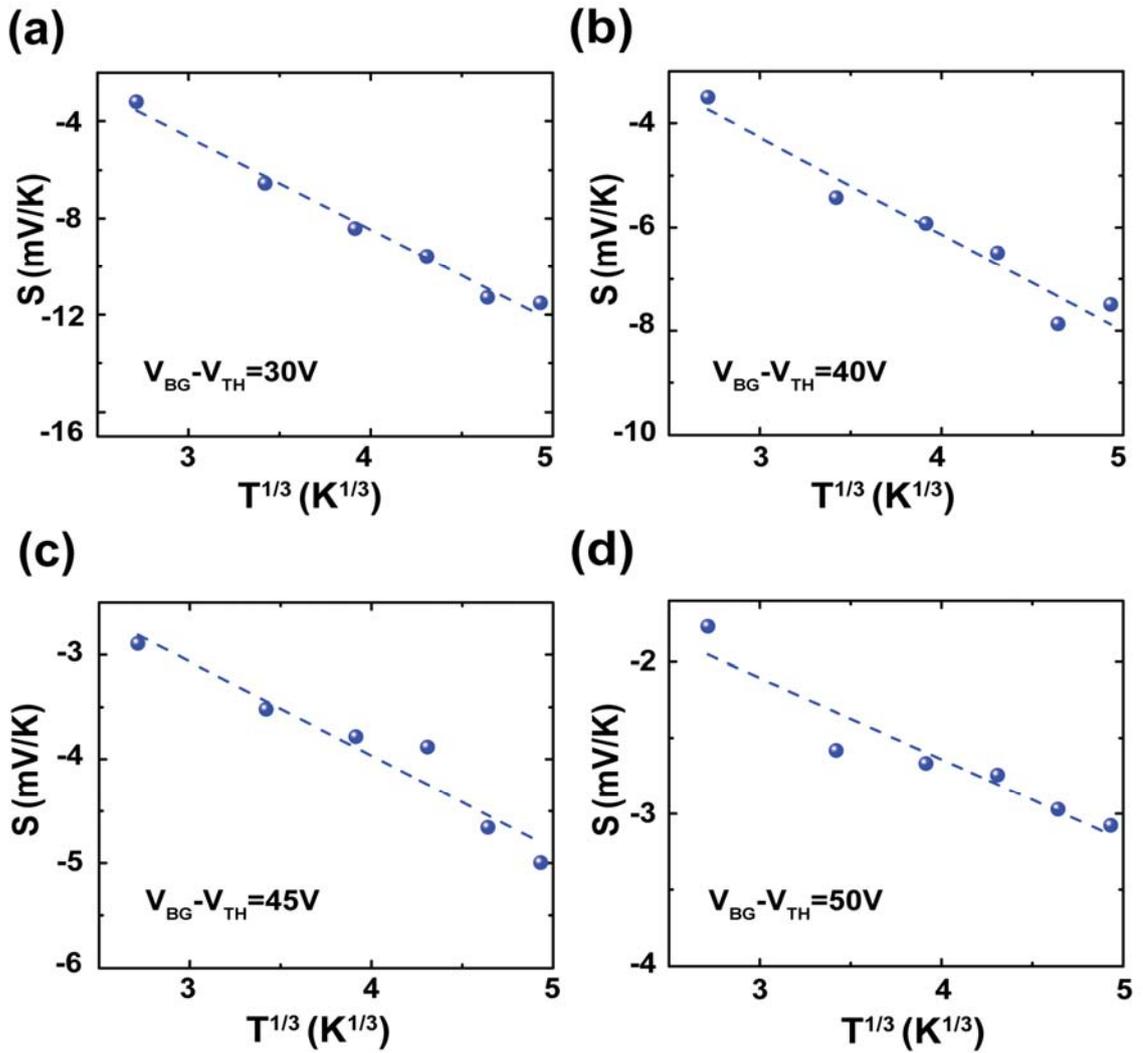

**Figure S9.** (a), (b), (c) and (d) show the TEP at the gate voltage $V_{BG}-V_{TH}$=30 V, 40 V, 45 V and 50 V.

TEP measurement of another sample is shown in Figure S10, being consistent with the results discussed in the main text. In this sample, we observed the similar shift on TEP when changing the environment temperature (Figure S10 (a)). The maximum TEP at room temperature is around 20 mV/K which also comparable to the previous sample. The link between conductance and the TEP (Figure S10 (b)) is similar as the one discussed in the main text.

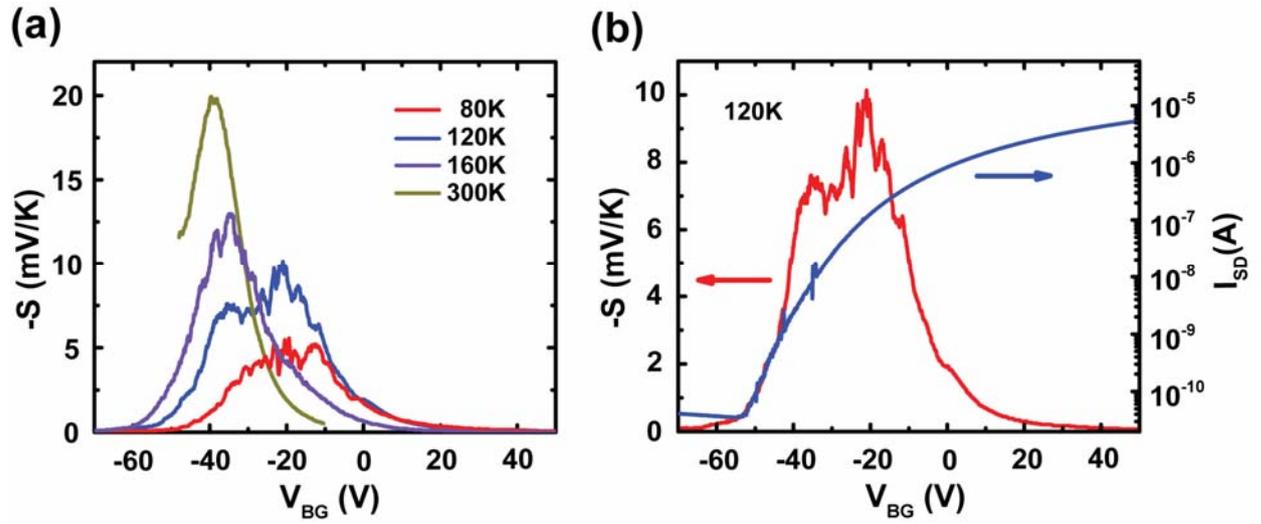

**Figure S10.** (a) TEP vs $V_{BG}$ at different temperatures. (b) Comparison of the gate dependent TEP and conductance at 120K.